 \documentclass[12pt,preprint]{aastex}

\usepackage{graphicx}				         
\usepackage{amsmath}
\usepackage{enumitem}			          	
\usepackage{amssymb}
\usepackage{pdfpages}

\newcommand {\OVII}   {\ion{O}{7}}
\newcommand {\OVIII}  {\ion{O}{8}}

\newcommand {\kms}    {km~s$^{-1}$}

\newcommand {\etal}   {et~al.} 
\newcommand{\msun}{$\rm{M}_{\odot}$}

\begin{document}
\title{Galactopause Formation and Gas Precipitation \\ 
During Strong Galactic Outflows}
\author{J. Michael Shull and Jacob A. Moss}
\affil {CASA, Department of Astrophysical \& Planetary Sciences. \\
University of Colorado, Boulder, CO 80309}

\email{michael.shull@colorado.edu, jacob.a.moss@colorado.edu}

 \date{}							


\begin{abstract}

Using X-ray constrained $\beta$-models for the radial distribution of gas in the outskirts of 
galaxies, we analyze the termination of galactic winds and the formation and evolution of 
halo clouds by thermal instability.  At low mass-loss rates, galactic winds are trapped within 
the halo, but they burst into the intergalactic medium during intermittent strong outflows with 
$\dot{M}_w \gg 1~M_{\odot}~{\rm yr}^{-1}$.   We develop analytic models of halo clouds as 
they cool radiatively over condensation time scales 
$t_{\rm c} \approx (390~{\rm Myr}) (T_6 /n_{-4}) (Z/Z_{\odot})^{-1}$ for hydrogen number 
densities $n_H \approx (10^{-4}~{\rm cm}^{-3})n_{-4}$, gas temperatures 
$T \approx (10^6~{\rm K})T_6$, and metallicities ($Z/Z_{\odot}$).  Halo gas can form 
kpc-scale clouds out to galactocentric distances $r \approx 30-65$~kpc, where efficient 
radiative cooling from $10^6$~K  down to $T \approx 10^4$~K occurs at 
$Z  \geq 0.3 Z_{\odot}$ on time scales less than 1~Gyr.  
After condensing to column densities $N_H \geq 3.5 \times 10^{16}$ cm$^{-2}$, these clouds 
lose hydrostatic pressure support and fall inward on dynamical time scales $\sim200$~Myr.  
Our baseline analysis will be followed by numerical calculations to understand the governing 
principles of halo cloud formation and transport of gas to the galactic disk.

\vspace{5cm}

\end{abstract}

\section{Introduction}

The circumgalactic medium (CGM) and intergalactic medium (IGM) help regulate the evolution of our 
Galaxy through the infall of gas that fuels continued star formation in the disk.  Progress in 
understanding these processes was stimulated by UV spectroscopic studies (Tumlinson \etal\ 2013; 
Stocke \etal\ 2013; Keeney \etal\ 2017) with the Cosmic Origins Spectrograph (Green \etal\ 2012) on the
{\it Hubble Space Telescope}.   Quasar absorption lines originating in the halos of intervening galaxies
discovered large metal-enriched gas reservoirs at impact parameters extending to radial distances 
of 100-150~kpc.   These gas clouds are unlikely to be at rest with respect to hot gas at the virial temperature
of the extended halo, as they are influenced by gravitational forces and intermittent pressure from galactic 
outflows.  After losing hydrostatic support,  halo gas falls inward as ``precipitation" to the galactic disk
(Voit \etal\ 2019) and observed as high-velocity clouds (HVCs) in the low halo (Wakker \etal\ 1999).  
The estimated infall rate of HVCs onto the disk (Shull \etal\ 2009; Putman \etal\ 2012; Fox \etal\ 2019) 
makes a significant contribution to the observed star formation rate (SFR) of 
$1-3~M_{\odot}~{\rm yr}^{-1}$ (Chomiuk \& Povich 2011).  

A key issue is characterizing the density perturbations that trigger condensations, which cool, condense, 
and fall to the disk.  These perturbations may arise from gas compression from galactic outflows and passing 
satellite galaxies.  Stalled galactic winds may produce a ``galactopause" (Shull 2014), a stagnation point for 
outflows marking a boundary between the CGM and IGM.  As we will show, the existence of a galactopause 
depends sensitively on the radial density structure of the halo gas.  Halos with density profiles decreasing more 
rapidly with radius than $n(r) \propto r^{-2}$ ($\beta > 2/3$ in the standard $\beta$-model) have  insufficient 
pressure to confine the outflows.  

In this paper, we present analytic formulations of thermal and dynamical processes in the extended halos 
of galaxies.   We explore the structure of a possible galactopause, at which gas outflows stall against the 
pressure of the CGM and IGM.  Accumulation of gas at this interface defines a sphere of chemical enrichment 
of the CGM.  During periods of starburst activity, galactic outflows can burst through the CGM and inject 
chemically enriched gas into the IGM.  In Section~2, we use gas-density models ($\beta$-models) to 
describe the CGM confining pressure and the extent of galactic winds.  The wind-CGM interface may furnish 
seeds for cloud formation, depending on the amount of compression and its range of influence. In Section~3,
we develop analytic descriptions of cloud formation by radiative cooling, thermal instability, cloud precipitation, 
and cloud interactions with the wind.   Section~4 concludes by describing effects on the long-term behavior of 
the CGM and implications for future research.

\section{Spatial Extent of Galaxy Halos}

In this section, we discuss the radial extent of galaxies.  As discussed previously (Shull 2014), galaxies may 
end by gravity or by gas outflows.  For gravitational estimates, we consider the virial radius ($R_{\rm vir}$) 
computed from the turnaround and gravitational collapse of over-dense perturbations in an expanding universe
and the gravitational radius ($r_g$) computed from the potential energy of spherical systems.   We then explore 
the confinement of galactic outflows by the pressure of hot gas in the extended halo.  The CGM can stall galactic 
winds and produce a galactopause, much like the heliopause that terminates the solar wind, and it marks the 
boundary with the interstellar medium.   

\subsection{Gravitational Extent} 

In their study of collapsed halos in cosmological N-body simulations, Navarro, Frenk, \& White (1997) found
that dark matter collapses into structures with cuspy cores and extended halos, fitted to radial (NFW) profiles
of density, potential, and enclosed mass, 
 \begin{eqnarray}   
  \rho(r)  &=&   \frac {\rho_0} {(r / r_s )  \left[ 1 +  (r / r_s ) \right] ^2  }    \\   
   \Phi(r) &=& - \left( 4 \pi G \rho_0 r_s^2 \right) \frac { \ln (1 + r/r_s)} {r/r_s}   \\
       M(r) &=& \left( 4 \pi \rho_0 r_s^3 \right) \left[ \ln \left( 1 +  \frac{r}{r_s} \right) - \frac {r / r_s}{1+ r / r_s} \right] \; .
\end{eqnarray}
Here,  $r_s = R_{\rm vir}/c$, is a characteristic radius related to the virial radius by a concentration parameter 
$(c)$.  For extended self-gravitating systems without a sharp boundary,  it is useful to define the ``gravitational 
radius" (Binney \& Tremaine 2008) as $r_g = GM^2/ \vert W \vert$, where the integrated gravitational 
potential energy is
\begin{equation}
    W = \frac{1}{2} \int  \,   d^3 \mathbf{x} \, \rho(\mathbf{x}) \, \Phi(\mathbf{x})  
        =  -4 \pi G \int_{0}^{\infty} \rho(r) \, M(r) \, r \, dr   \;  .  
 \end{equation}
Although the NFW mass diverges logarithmically with radius, the gravitational radius can be found by 
integrating eq.\  (4) out to $R_{\rm vir} = c r_s$,
\begin{eqnarray}
   W_{\rm vir} ^{\rm (NFW)}  &=& - \left( \frac  {GM_{\rm vir}^2} {2 \, r _s} \right) 
    \frac { \left[ 1 - \frac {\ln (1+c)} {(1+c)}  - \frac {1}{(1+c)} \right] }
   { \left[  \ln (1+c) - \frac {c}{1+c} \right]^2  }    \\
 r_g ^{\rm (NFW)}             &=&  GM_{\rm vir}^2 / \vert W_{\rm vir} \vert = 
         (2 r_s) \;  \frac {  \left[ \ln (1+c) - \frac {c}{1+c} \right]^2  } 
          {  \left[ 1 - \frac {\ln (1+c)} {(1+c)}  - \frac {1}{(1+c)} \right] }  \; .
\end{eqnarray}   
The gravitational radius is similar to the ``half-light radius", $r_h  \approx 0.45 (GM^2/ \vert W \vert)$,
defined by Spitzer (1969) for spherical stellar systems with various mass distributions.  For most 
NFW halos, the ratio of gravitational radius to virial radius, $(r_g/R_{\rm vir}) = (r_g/ cr_s)$, is 0.5-0.7.   
For example,  $r_g = 3.44r_s = 0.69 R_{\rm vir}$ for $c=5$, 
and $r_g = 10.8 r_s = 0.54R_{\rm vir}$ for $c = 20$.  

The virial radius is commonly used as an estimate of the collapsed region around halos of dark matter,
including galaxies, groups, and clusters of galaxies.  The original definition adopted critical mass 
overdensities $\Delta_{\rm vir} \approx 200$ times the ambient density at collapse. That formulation 
was based on an outdated cosmological model with closure density in matter ($\Omega_m = 1$).  
Using criteria for galaxy collapse and mass assembly in a $\Lambda$CDM universe, Shull (2014) 
defined the virial radius,
\begin{equation}
   R_{\rm vir} (M_h, z_a) = (206~{\rm kpc}) \, h_{70}^{-2/3} \, M_{12}^{1/3}  \left[  \frac 
       {\Omega_m (z_{\rm a})  \, \Delta_{\rm vir} (z_{\rm a})}  {200} \right] ^{-1/3} 
        \left( 1+z_a \right)^{-1}  \;  ,
\end{equation}
for a galaxy associated with total halo mass $M_h = (10^{12} M_{\odot}) M_{12}$.  The extra factor
$(1+z_a)^{-1}$ appears because virialization is assumed to occur at the assembly redshift ($z_a$) 
when the background density was higher by a factor $(1+z_a)^3$.  This redshift was determined 
from cosmological collapse criteria (Lacey \& Cole 1993;  Sheth \& Tormen 1999).  
As computed by Trenti \etal\ (2013), these virialization redshifts range from $z_a \approx 1.35$ 
for $M_h = 10^{11}~M_{\odot}$ to $z_a \approx 0.81$ for $M_h = 10^{14}~M_{\odot}$.  

\subsection{Wind Confinement and Galactopause} 

An estimate for the termination radius for a galactic wind in a biconical outflow into solid angle 
$\Omega_w$ for constant CGM thermal pressure  $P_{\rm CGM} = n kT_{\rm CGM}$ is,
\begin{equation}
      R_{\rm term} = \left( \frac { \dot{M}_w \, V_w } { P_{\rm CGM} \Omega_w } \right)^{1/2}
               \approx (138~{\rm kpc} ) \left[ \frac { \dot{M}_{10} \,  V_{200} } {P_{40} \,  (\Omega_w /4 \pi) }
                   \right]^{1/2}  \;  .
 \end{equation}
This radius marks the location where the wind ram pressure, $P_{\rm w} = \rho_w  V_w^2$, 
in a steady-state mass outflow rate, $\dot{M}_w = \Omega_{\rm w} r^2 {\rho_{\rm w} V_{\rm w}}$, 
stagnates against the thermal pressure.  The mass-loss rate is scaled to
$\dot{M}_w = (10~M_{\odot}~{\rm yr}^{-1}) \dot{M}_{10}$ with wind speed 
$V_w = (200~{\rm km~s}^{-1})V_{200}$.  We scale the solid-angle coverage of the outflow to 
$(\Omega_w / 4 \pi)$ and the confining thermal pressure to 
$P_{\rm CGM} /k = (40~{\rm cm}^{-3}~{\rm K}) P_{40}$.  The total particle number density is 
$n = n_{\rm H} + n_{\rm He} + n_e \approx 2.247 n_{\rm H}$ for fully ionized gas 
(H$^+$, He$^{+2}$, $e^-$) with $n_e = n_{\rm H} + 2 n_{\rm He} = 1.165 n_H$.  
We adopt $Y = 0.2477$ for the He/H ratio by mass and $y = 0.0823$ by number (Peimbert \etal\ 
2007), comparable to estimates of the primordial value, $Y_p = 0.2449 \pm 0.0040$ (Cyburt \etal\ 
2016).  Typical galactic outflows have $\dot{M}_w = 1-10~M_{\odot}~{\rm yr}^{-1}$,  
$V_w = 100-300$~\kms,  and  $\Omega_w / 4 \pi = 0.2-0.4$ (Veilleux \etal\ 2005).  Inferred gas
pressures in the CGM range from $P/k = 10-50~{\rm cm}^{-3}$~K (Keeney \etal\ 2017), appropriate 
for halo virial temperature $T_{\rm CGM} \approx 2 \times 10^6$~K and hydrogen number density  
$n_H \approx 10^{-5}~{\rm cm}^{-3}$ at $r \approx 100$~kpc.
The outflow rate can be related to the SFR by a mass-loading factor $\beta_m = \dot{M} / {\rm SFR}$
inferred from optical and X-ray data to lie in the range $\beta_m = 1-3$ (Strickland \& Heckman 2009).   

For these parameters, we estimate that most winds terminate at 
$R_{\rm term} \approx 100-200$ kpc during stages of moderate star formation. Some of the gas in 
the winds escapes to the IGM, and a portion is recycled back to the disk on a free-fall timescale 
$t_{\rm ff} \approx 1~{\rm Gyr}$.  Stronger winds concentrated into biconical outflows with 
$\Omega_w \approx 0.3 \times 4\pi$ shift the termination radius to much larger radii, where 
these assumptions break down.  The estimate in equation (8) is also unreliable because it assumes 
a constant confining pressure from a hot CGM.   More realistic estimates for wind termination must 
account for the decrease in density with radius observed in the outskirts of galaxies 
(Bregman \etal\ 2018).

For a better analysis of wind termination and its sensitivity to the CGM, we employ the  ``$\beta$ model"
(Cavaliere \& Fusco-Femiano 1976; Sarazin 1986) often used to estimate the density in the outskirts of 
galaxies and groups.   For the Milky Way halo, these models are constrained by X-ray emission
lines and absorption lines (e.g., \OVII\ and \OVIII).  X-ray surface brightness profiles have also been
fitted for some external galaxies, with considerable uncertainty in their outer regions.  Following the 
approximation adopted by Miller \& Bregman (2013), we write the number density as
\begin{equation}
      n(r) = n_0  \left[  1+ \frac {r^2} {r_c^2} \right]^{-3\beta / 2} \approx [n_0 r_c^{3 \beta}] \, r^{-3 \beta} \; .
\end{equation}
The latter approximation is valid in the limit of $r \gg r_c$, where $r_c$ is the core radius and 
$n_0$ is a fiducial number density\footnote{The $\beta$-models developed by Miller \& Bregman 
(2013, 2015) refer to $n(r)$ as ``the gas density", presumably the electron density $n_e(r)$, as
noted explicitly in Voit  (2019).  We assume fully ionized gas with He/H = 0.0823, so that 
$n_e = 1.165 n_H$.  The total particle density is then $n = 2.247 n_H = 1.929 n_e$.}
Miller \& Bregman (2013) found that a spherically symmetric $\beta$ model works just as well as a 
variation that considers the geometry of the galaxy, and we use it for simplicity.  The galactopause 
lies well outside of the core radius, so that the wind typically has a termination shock in the CGM.   
For specific calculations, we adopt the parameters of Bregman \etal\ (2018), with index 
$\beta = 0.51 \pm 0.02$ and normalization 
$[n_0 r_c^{3 \beta}] = (2.82 \pm 0.33)\times10^{-2}~{\rm cm}^{-3}~{\rm kpc}^{3 \beta}$, where 
$r_c$ is in kpc units.   The wind outflow stalls against CGM thermal pressure in a halo described   
by a $\beta$-model  when
\begin{equation}
   \frac  { 1.93 \, [n_0 r_c^{3 \beta}] \; kT} { r^{3 \beta} } =   \frac { \dot {M_w} V_w} { \Omega_w \; r^2 } \; .
\end{equation}
Adopting $\beta = 0.51$ and scaling the mass-outflow rate of strong winds as 
$\dot{M}_w = (10~M_{\odot}~{\rm yr}^{-1}) \dot{M}_{10}$, we find that  the galactopause (GP) 
occurs at radius
\begin{equation}
   R_{\rm GP} = \left[ \frac { \dot {M}_w V_w / \Omega_w kT } {1.93 (n_0 r_c^{3 \beta})
         (3.086\times10^{21}~{\rm cm})^2 } \right] ^ {1/(2-3\beta)} =
       (277~{\rm kpc}) \left[ \frac {\dot {M}_{10} V_{200} } {T_6 (\Omega_w/4 \pi) } \right] ^{2.13}   \; .
\end{equation} 
We could also include the small decline in temperature with radius expected for halos near hydrostatic 
equilibrium with lower circular velocities in their outer regions.   In hydrostatic equilibrium, 
$d \ln P / d \ln r = -2 T_{\phi} / T$, where the ``potential temperature" $T_{\phi} \equiv (\mu V_c^2/2k)$
assuming thermal energy of $kT/2$ for each of the two kinetic-energy components of circular velocity
$V_c$.   At the solar distance in the Milky Way, $V_c \approx 230~{\rm km~s}^{-1}$ and 
$T_{\phi} \approx 1.9 \times 10^6$~K for mean particle mass $\mu = 0.592m_H$.  
 The circular velocity peaks at $r  = 2.163 r_s$ for NFW halos ($0.216 R_{\rm vir}$ for $c = 10$) and 
decreases slowly at larger distances into the halo.  Models that adopt a constant ratio of cooling time  
to freefall time (Voit 2019) find shallow temperature profiles with $T(r) \propto r^{-0.1}$.  

The expression above illustrates the sensitivity of the potential galactopause to the radial density 
profile.  Because of the close competition between decreasing wind ram pressure
($P_{\rm ram} \propto r^{-2}$) and CGM thermal pressure ($P_{\rm CGM} \propto r^{-1.53}$), the 
galactopause radius scales non-linearly with parameters of the outflow strength (mass-loss rate, 
wind velocity, solid angle of the outflow).  Because each of these quantities varies among galaxies, 
and will change over time in any individual galaxy, the wind termination moves inward and 
outward in radius.  For example, a doubling of the outflow strength, either $\dot{M}_w$ or $V_w$, 
would increase $R_{\rm GP}$ by over a factor of four, resulting in the wind bursting out of the CGM.
Similarly, a decrease in the outflow strength would trap the wind, with a
termination  shock at $r < 100$~kpc.  

In equation (11), the wind mass-loss rate ($\dot{M}_w$) was expressed in units of 
$10M_{\odot}~{\rm yr}^{-1}$ flowing into total solid angle $\Omega_w$.   The CGM temperature 
was scaled to $T_{\rm CGM} = (10^6~{\rm K}) T_6$.   These winds will have sufficient momentum
to break through the CGM when $R_{\rm GP} \geq R_{\rm CGM}$ at
\begin{equation}
   \dot{M}_w \geq (10.4~M_{\odot}~{\rm yr}^{-1}) \left( \frac {R_{\rm CGM} } {300~{\rm kpc} } \right)^{0.47}
          T_6 \; V_{200}^{-1} \left( \frac {\Omega_w}{4 \pi} \right)   \; .
\end{equation}
This break-through criterion is sensitive to $\beta$ through the radial gradient of $n(r) \propto r^{-3 \beta}$, 
particularly for values of $\beta > 0.5$.  Previous studies of \OVII\ absorption in the Milky Way halo found 
$\beta = 0.56^{+0.10}_{-0.12}$ (Miller \& Bregman 2013),  $\beta = 0.53 \pm 0.03$ (Hodges-Kluck \etal\ 
2016), and $\beta = 0.51 \pm 0.02$ (Bregman \etal\ 2018).   Fits to \OVII\ and \OVIII\ emission lines in
the Milky Way halo gave $\beta = 0.50 \pm 0.03$ (Miller \& Bregman 2015).  These density profiles are 
all sufficiently steep ($r^{-1.5}$ to $r^{-1.7}$) that strong outflows will break out of the CGM.   However, 
X-ray observations in the outskirts of galaxies remain uncertain. 

\section{Halo Cloud Evolution with Radiative Cooling}

\subsection{Thermal Instability and Cloud Compression}

We now discuss cloud evolution during the cooling phase and the possible effects of stripping and 
re-assembly of the detached parcels of gas.    Thermal instability followed by precipitation 
(Voit \etal\ 2019) is a plausible method of forming cool gas clouds in galaxy halos at a speed governed 
by the radiative cooling rate per unit volume, $n_e n_H \Lambda(T)$.   Over the temperature range 
$5.0 < \log T < 6.5$, we approximate the cooling function (Gnat \& Ferland 2012) as 
$\Lambda(T) \approx \Lambda_0 \, (T/T_0)^{-0.7}$,  with 
$\Lambda_0 = (2 \times 10^{-22}~{\rm erg~cm}^{3}~{\rm s}^{-1})(Z/Z_{\odot})$ at fiducial temperature 
$T_0 =(10^6~{\rm K})T_6$.  The linear scaling of $\Lambda$ with metallicity is valid for 
$0.1 < (Z/Z_{\odot}) < 1$.  For fully ionized gas in the galactic halo, this formula leads to an {\it initial} 
cooling time and cooling length for typical conditions at  radial distances $r \approx 50$~kpc, 
\begin{eqnarray}
   t_{\rm cool} &=& \frac{3nkT/2} {n_e n_H \Lambda(T)} \approx (630~{\rm Myr}) \, T_6^{1.7}
                                           \left( Z / Z_{\odot} \right)^{-1} n_{-4}^{-1}   \\
   \ell_{\rm cool} &=& c_s(T) \,  t_{\rm cool} \approx (98~{\rm kpc}) \, T_6^{2.2}
                                           \left( Z / Z_{\odot} \right)^{-1} n_{-4}^{-1}   \; .
 \end{eqnarray}  
We evaluated these quantities for hydrogen number density $n_H = (10^{-4}~{\rm cm}^{-3}) n_{-4}$, 
with $n_{\rm He} = 0.0823 n_H$, $n_e = 1.165 n_H$, $n = 2.247n_H$, $\mu = 0.592m_H$, and 
adiabatic sound speed $c_s  = (5kT/3\mu)^{1/2} \approx (152~{\rm km~s}^{-1}) T_6^{1/2}$.
This illustrates the dramatic decrease of $\ell_{\rm cool}$ when hot halo gas is triggered into rapid
cooling at constant pressure.   As the gas cools below the peak of the cooling function, at 
$T \approx 10^{5}$~K,  the timescales drop to much smaller values, $t_{\rm cool} \approx 10$~Myr 
and $\ell_{\rm cool} \approx 1$~kpc.  The non-linear dependence of these parameters on temperature 
produces rapid cloud evolution.

To describe the basic effects of cooling on cloud compression, we develop an analytic model of a 
cooling spherical cloud of radius $R$ and temperature $T$.  We make a few simplifying assumptions: 
mass conservation ($nR^3 =$ constant), pressure equilibrium ($nT =$ constant), and enthalpy change 
with compression and radiative cooling.  The cloud compresses with $n\sim T^{-1}\sim R^{-3}$, and the 
gas radiates away energy at constant pressure, with a cooling function 
$\Lambda(T) = \Lambda_0 (T/T_0)^{-0.7}$ and change in enthalpy over cloud volume 
$V_{\rm cl} = (4 \pi R^3/3)$ is 
\begin{equation}
     \frac{dH}{dt} = -n_en_{\rm{H}}\Lambda(T) \, V_{\rm cl} 
                = \frac{d}{dt}\left[\frac{5}{2}nkT \times  \frac{4 \pi}{3} R^3\right].
\end{equation}
Approximating $R^{-7.1}\approx R^{-7}$ leads to a simple differential equation,
\begin{equation}
R^7 \, \frac{dR}{dt}=-\left[\frac{\Lambda_0n_0^2R_0^8}{14.38P_0}\right]  \;,
\end{equation}
with analytic solutions,
\begin{eqnarray} 
R(t) &=& R_0 \left[ 1- (t/t_{\rm c}) \right]^{1/8}  \\
T(t ) &=& T_0 \left[ 1-(t/t_{\rm c}) \right]^{3/8}   \\
n(t) &=& n_0\left[ 1-(t/t_{\rm c}) \right]^{-3/8}    \; ,
\end{eqnarray}
for initial conditions in radius ($R_0$) and pressure ($P_0 = n_0kT_0$).  
These relations are valid as the cloud cools from $10^6$~K to $10^5$~K, after a time 
$t \approx 0.998 t_{\rm c}$ and at radius $R \approx 0.464 R_0$.  Once $T < 10^5$~K, $\Lambda(T)$
drops below its peak rate.   Cooling and compression continue, as the cloud approaches photoionization 
equilibrium with the ionizing background at temperature $T \approx10^4$~K.  The total time-scale for 
clouds to cool below $10^5$~K and condense sufficiently to lose pressure support is
\begin{equation}
       t_{\rm c}=1.80  \left( \frac {kT_0} { n_0 \Lambda_0} \right)  \approx (390~{\rm Myr}) 
                (T_6 / n_{-4}) \, (Z/Z_{\odot})^{-1}  \; .
\end{equation}
 
 Cloud condensation is triggered by initial conditions sufficient to cool the halo gas on time scales
  $t_{\rm cool} \leq1$~Gyr.  Gas densities in the halos of galaxies are somewhat uncertain, with 
 estimates  ranging from  $n_H \approx 10^{-5}$ to $10^{-4}~{\rm cm}^{-3}$ (Voit 2019).  For the 
 Milky Way, the hydrogen number density at $r = 50$~kpc has been estimated at 
 $\sim10^{-4}~{\rm cm}^{-3}$ (Salem \etal\ 2015; Faerman \etal\ 2017; Bregman \etal\ 2018).  
 Figure 3 of Voit (2019) shows a similar range of density profiles in the Milky Way halo, based on 
 multiple observational methods.  We therefore adopt a halo model, scaled to typical conditions at 
 $r_0 = 50$~kpc, with $n_{\rm H} \approx (1 \times 10^{-4}~{\rm cm}^{-3})(r/r_0)^{-1.5}$.   Gas compression 
 could be triggered by the injection of gas from the disk plane,  gas stripped from infalling satellite 
 galaxies, or by stalling of winds at the galactopause.   These clouds will cool, condense, lose hydrostatic 
 pressure support, and fall toward the disk.  From equation (13) we see that cloud formation can occur 
 on initial cooling time scales $t_{\rm cool} < 1$~Gyr out  to distances,
\begin{equation}
     r_{\rm cool} \leq (30~{\rm kpc}) \, T_6^{-1.13} \left( \frac {Z}{0.3Z_{\odot}} \right)^{2/3}  \; , 
\end{equation}  
 scaled to sub-solar halo metallicities ($Z \approx 0.3 Z_{\odot}$) and temperatures $\sim10^6$~K.    
 If temperatures in the outer halo decrease with the declining potential temperature,
 $T_{\phi} = (\mu V_c^2/2k) \approx 5\times10^5$~K, cooling times would shorten, allowing clouds 
 to form out to $r \approx 65$~kpc.   The average CGM density profile will likely not form kpc-scale 
 clouds beyond these distances.  However, the outer formation radius 
 could be larger in the presence of compressional triggers.  


\subsection{Cloud precipitation}

Newly formed clouds will contract while cooling, if they are in pressure equilibrium with the surrounding 
gas. This decreases the difference in pressure between the top and bottom of the cloud, causing the 
cloud to fall out of hydrostatic equilibrium with the hot ambient halo gas. To estimate when a cloud loses 
pressure support in the halo, we find where the force of gravity on the cloud is greater than the pressure 
difference between the top and the bottom of the cloud, $M_{\rm slab} \: g(z) > (P_1-P_2) A$.  
We approximate the cloud as a slab with cross-sectional area $A$, vertical extent $\Delta z$, and mass 
$M_{\rm slab}= 1.33 m_H n_H A (\Delta z)$ for $(\rm{He/H})=0.0823$ by number.  Rewriting the total 
hydrogen column density, $N_H=n_H (\Delta z)$, we express the criterion for cloud precipitation as
$N_H > (P_1-P_2) / 1.33 m_H \, g(z)$, with $P_1$ defined as pressure at the lower side of the cloud.  
We assume that the enclosed galactic mass is dominated by a spherically symmetric DM halo, with 
$g(r) = GM(r)/r^2$ for $r \gg r_{\rm disk}$.  For an isothermal medium with $T = 2\times10^6$~K, 
ambient pressure differences are due to density gradients in the CGM.  We approximate the total particle 
density as $n(r) = n_0 (r/r_0)^{-3\beta} $, with $r_0 = 50$ kpc and $\beta = 0.51$.  We assume that the 
cloud lies near the polar axis, with $n(z) = n(r)$, $g(z) \approx g(r)$, and $\Delta P = kT (\Delta n)$, 
\begin{equation}
      \Delta n = \Bigl| \frac {dn} {dr} \Bigr| \Delta z = \left (\frac {3\beta n_0} {r_0} \right) 
             \left( \frac {r} {r_0} \right)^{-(3\beta+1)} \Delta z.
\end{equation}
For a flat rotation curve, we approximate the enclosed mass $M(r) = M_0(r/r_0)$ with $r_0 = 50$ kpc.
Recent estimates for the Milky Way (Callingham \etal~2019) find a virial mass 
$M_{\rm vir}=1.17 \times 10^{12}$~\msun\ and $M(r\leq20~{\rm kpc}) = 0.12 M_{\rm vir}$.
For an NFW model with $M_{\rm vir} = (4\pi\rho_0r^3_s) \left[ \ln(1+c)- c/(1+c) \right]$ and $c = 10$, 
the enclosed mass within 50~kpc is $M_0 = 3.89 \times10^{11}~M_{\odot}$, comparable to the value,
$4 \times10^{11}~M_{\odot}$ found by Deason \etal\ (2012).  With mean cloud pressure 
$P_0 = 2.247n_0 kT_0$ at $r_0$ and slab thickness $\Delta z=1$~kpc, the minimum 
column density for precipitation is
\begin{equation}
       N_H > \frac {3\beta n_0 kT_0} {1.33m_H  r_0} \left( \frac {r} {r_0} \right)^{-2.53}
          \frac {\Delta z}{g(r)}   
             \approx (3.5 \times 10^{16}~{\rm cm}^{-2}) \left( \frac {r} {50~{\rm kpc} } \right)^{-1.53} 
                \left[ \frac {P_0/k} {40~\rm{cm}^{-3}~K} \right]    \; .
\end{equation}
This column density is similar to that in the lowest column density HVCs (Collins \etal\ 2003)
which can extend up to $N_H \sim 10^{20}~{\rm cm}^{-2}$ (Wakker \etal\ 1999).  Photoionization 
models of HVCs with $R = 0.5-2$~kpc and $n_{\rm cl}=10^{-3}$ to $10^{-2}~{\rm cm}^{-3}$ typically 
have total hydrogen column densities of $10^{18-20}~{\rm cm}^{-2}$.

\newpage

\subsection{Cloud Infall and Stripping}

When a thermally unstable cloud with sufficient column density loses pressure support, it begins falling 
inward, a process referred to as ``precipitation" (Voit \etal\ 2019).  For standard dark-matter halos containing 
a CGM with hot gas, we can estimate the cloud dynamics with a simple model of  a 1 kpc cloud with total 
particle density $n_{cl}=10^{-3}~\rm{cm}^{-3}$ at initial distance $r_0 = (50~{\rm kpc})r_{50}$ from the
galactic center on the polar axis.  In free fall,  its inward acceleration is $g(r) \approx -GM(r)/r^2$, neglecting 
hydrostatic pressure gradients and cloud drag forces (which we consider later).  We can estimate the cloud
infall velocity in the halo, over the region with a flat rotation curve where $M(r) = M_0(r/r_0)$, using the first 
integral of motion with initial conditions $\dot{r}=0$ and $r = r_0$ at $t  = 0$, 
\begin{equation}
     \frac {1}{2} \dot{r}^2 =-\int_{r_0}^{r}\frac{GM(r)} {r^2}dr = -\int_{r_0}^{r}\frac{GM_0/r_0} {r}dr \; . 
\end{equation}
The cloud infall velocity has a weak dependence on radius,
\begin{equation}
      \dot{r}=-\left[\frac{2GM_0}{r_o}\right]^{1/2}\sqrt{\ln(r_0/r)} \; .
\end{equation}
With the substitution $r = r_0 \exp(-u^2)$, the drag-free infall can be described as
\begin{equation}
    t (r) = \left(  \frac {r_0^3}{2GM_0} \right)^{1/2} \int_0^{u(r)} \exp(-u^2) \, du = 
                \left(  \frac {\pi r_0^3}{8GM_0} \right)^{1/2}  {\rm erf} \,  [u(r)]   \; ,
\end{equation}
where $u(r) = [ \ln (r_0/r)]^{1/2}$ is the dimensionless argument of the Gaussian error function.   
We define a dynamical infall time in the dark-matter halo, from $r_0$ to $r \ll r_0$ where the error 
function approaches 1.  For the Milky Way, with $M_0 = 3.89\times10^{11}~M_{\odot}$ interior to
50~kpc, the infall time and characteristic velocity are,
\begin{equation}
   t_{\rm in} \approx \left[  \frac {\pi r_0^3 }{8GM_0} \right]^{1/2} \approx (190~{\rm Myr})  r_{50}^{3/2}  
       \; \; \;  {\rm and} \; \; \; (2GM_0/r_0)^{1/2} \approx (260~{\rm km~s}^{-1}) r_{50}^{-1/2} \; .  
\end{equation}

As noted previously, the density perturbations from thermal instability in diffuse halo gas do not remain 
stationary relative to the hot halo gas.  Clouds in the halo are acted upon by gravitational forces and 
ram-pressure stripping;  during periods of  active star formation they can be buffeted by galactic winds.  
The clouds typically achieve velocities of 100-300 \kms\ relative to hot halo gas, often assumed to be in 
hydrostatic equilibrium.  For  cloud infall through a static halo with no wind, stripping begins when the 
ram pressure $\rho_{\rm halo} \dot{r}^2$ exceeds the internal cloud pressure, $n_{\rm cl} kT_{\rm cl}$.  
We assume gas mass density $\rho_{\rm halo} = 1.33 m_H n_H(r)$ with $n_H(r) = n_{H,0} (r/r_0)^{-\beta}$.   
From Equation (25), cloud stripping starts when
\begin{equation}
    \left(  r_0 / r \right) ^{3 \beta} \ln (r_0 / r) \geq \left[ \frac {P_{\rm cl}}{1.33 \, m_H \, n_{H,0}} \right] 
                    \left( \frac {2GM_0}{r_0} \right)^{-1}   \; .
\end{equation}
For a condensing cloud with $P_{\rm cl}/k \approx 225~{\rm cm}^{-3}~{\rm K}$ 
($n_H = 10^{-3}$~cm$^{-3}$, $T = 10^4$~K) in a halo with $n_{H,0} = 10^{-4}$~cm$^{-3}$ and
$\beta = 0.51$, the right-hand side of equation (28) equals 0.209.  Stripping begins when the cloud
falls to a fraction $(r/r_0) \approx 0.85$ of its initial radius.  

During periods of active star formation, galactic winds sweep over newly condensed halo clouds.  The gas 
dynamics of cloud-wind interaction has been studied numerically by many authors (Stone \& Norman 1992;
Mac Low \etal\ 1994; Klein \etal\ 1994; Orlando \etal\ 2005; Silvia \etal\ 2010).  In the wind-cloud encounter, 
we assume that the momentum imparted to the cloud depends on the fraction of wind mass striking the 
geometric cross section, with radius $R = (500~{\rm pc})R_{500}$.  We adopt a density contrast  
$\Delta_{\rm cl} = (n_{\rm cl} / n_H) = 100 \Delta_{100}$ between the cloud and ambient halo gas, of 
density $n_{\rm H,0} = (10^{-4}~{\rm cm}^{-3})n_{-4}$ at $r = 50$~kpc.  When a wind with 
$V_w = (200~{\rm km~s}^{-1})V_{200}$ impacts the cloud, it sends a slow shock of velocity
$v_s = V_w \Delta^{-1/2} \approx (20~{\rm km~s}^{-1}) V_{200} \Delta_{100}^{-1/2}$ through the
cloud, transiting on a crossing time, 
\begin{equation}
     t_{\rm cr} \approx (2R/v_s) = (49~{\rm Myr}) R_{500} V_{200}^{-1} \Delta_{100}^{1/2}  \; .
\end{equation} 
Most simulations find that clouds are shredded on a times scale $\sim2t_{\rm cr}$.
For a cloud of mass $m_{\rm cl}$, the intercepted mass is the product of wind mass flux,  
cloud cross section, and cloud-crossing time,
\begin{eqnarray} 
  \Delta m    &=& (\dot{M}/\Omega_w \, r^2)(\pi R^2)(2R/v_s)  \approx 
       (1.2 \times 10^4~M_{\odot})   (4 \pi / \Omega_w) R_{500}^3 \dot{M}_{10} \, V_{200}^{-1} 
              \Delta_{100} ^{1/2}   \\  
  m_{\rm cl} &=& (4 \pi R^3/3) (1.33 n_H m_H \Delta_{\rm cl}) \approx
             (1.7 \times10^5~M_{\odot}) R_{500}^3 \, n_{-4} \, \Delta_{100} \; .
\end{eqnarray} 
Comparing $\Delta m$ to $m_{\rm cl}$, we see that halo clouds can intercepts a moderate amount 
of wind material.  The fractional mass gain is independent of cloud radius $R$, but it depends on
outflow parameters and CGM conditions at $r = 30-100$~kpc,
 \begin{equation}
    (\Delta m / m_{\rm cl}) \approx (0.072)  (4 \pi / \Omega_w) {\dot M}_{10} V_{200}^{-1} \, 
                n_{-4} \, r_{50}^{-2} \, \Delta_{100}^{-1/2} \; .  
 \end{equation}
 Momentum conservation during the wind-cloud encounter results in a shift in velocity,
 \begin{equation}
      \Delta V_{\rm cl} = (\Delta m/ m_{\rm cl}) V_w \approx (14~{\rm km~s}^{-1})
        (4 \pi / \Omega_w) {\dot M}_{10} \, n_{-4} \, r_{50}^{-2} \, \Delta_{100}^{-1/2} \; .
 \end{equation} 
Because of the range in outflow strengths ($\dot{M}$ and $\Omega_w$) and CGM density, 
this velocity impulse varies widely, from a few \kms\ to much larger values. Particularly for clouds 
at $r < 50$~kpc swept by biconical outflows with $(\Omega/4  \pi)  \approx 0.3$, winds can
significantly affect the structure of the cloud, sometimes shredding it.   However, some  parcels of stripped 
gas may re-accrete onto the cloud, if they experience less drag than the main cloud that precedes it.  
This is akin to  the ``Peloton effect" well known in bicycle racing and bird flying formation.

\section{Summary and Discussion}

The purpose of this paper was to investigate the formation and evolution of clouds in the galactic halo
as they cool, condense, and lose hydrostatic pressure support from the hot gas.  The CGM is not a 
passive region between the IGM and the galactic disk, but an environment conducive to forming 
clouds by thermal instability.   We developed a variety of analytic models to estimate the radial extent 
of large-scale galactic outflows and their influence on halo cloud formation and infall to galactic disks.  
Recent observations suggest that the CGM is enriched with heavy elements, consistent with injection
of disk gas into the halo through galactic winds during periods of active star formation.  After new clouds
form and condense, they fall inward on 200-Myr timescales, providing material for continued star formation
in the disk.  Cooling and compression of the clouds occurs on longer timescales, 
$t_{\rm c} \approx (390~{\rm Myr}) (T_6 /n_{-4}) (Z/Z_{\odot})^{-1}$, possibly triggered by external
compression events.   After condensing to column densities $N_H \geq 3.5 \times 10^{16}$ cm$^{-2}$, 
these clouds lose hydrostatic pressure support and fall inward.  

A critical parameter for wind termination in a galactopause is the radial pressure profile of the halo gas. 
Much of the outflow mixes with the CGM, leading to over-densities and nonlinear thermal instability.  
Observations show that galactic outflows are multi-phased, spanning temperatures from 
$10^8$~K to $10^2$~K (Zhang 2018).   Through thermal instability, cooling and compressed gas at 
the galactopause or in the diffuse CGM could provide seeds for cloud formation.  The CGM should 
constantly form rapidly cooling gas at $T  \leq10^5$~K.  The cooling parameters suggest that smaller 
clouds form more frequently at smaller galactic radii.   Cloud formation should occur out to radial 
distances $\sim 30-65$~kpc, depending on how steeply density, temperature, and metallicity fall off
with radius.   From our analytic models, the primary conclusions  are as follows.
\begin{enumerate}

\item For radial density profiles, $n(r) = n_0(r/r_0)^{-3\beta}$, with $\beta \approx 0.5$ and
$n_0 \approx 10^{-4}~{\rm cm}^{-3}$ at $r_0 \approx 50$~kpc, typical outflows will stall at 
galactocentric distances $r \approx 200-300$~kpc.  The galactopause depends on wind 
strength parameters ($\dot{M}_w$, $V_w$, $\Omega_w$).  If $\beta > 0.6$, the winds will have
no termination shock in the CGM.  Strong outflows with sufficiently fast winds will escape the 
galactic halo and burst into the IGM.

\item In strong galactic outflows ($\dot{M}_w > 10~M_{\odot}~{\rm yr}^{-1}$) wind material will
mix with the CGM and interact with existing clouds.   Depending on the timing, these winds 
could trigger cloud cooling through compression.

\item The cooling time $t_{\rm cool} \propto T_6^{1.7}$ and cooling length 
$\ell_{\rm cool} \propto T_6^{2.2}$ have strong temperature dependences for $10^7$~K down 
to $10^5$~K.  Metal-enriched gas that is slightly cooler than the virial temperature will cool 
and produce kpc-scale clouds in less than 1~Gyr.  

\item Clouds can form by thermal instability out to radii $r \approx 30-65$ kpc for halo density 
profiles with $\beta \approx 0.5$ and $n_H \approx 10^{-4}~{\rm cm}^{-3}$.  Efficient radiative 
cooling with $t_{\rm cool} < 1$~Gyr should occur at metallicities $Z  \geq 0.3 Z_{\odot}$.  

\item  Newly formed clouds that condense to $N_H \geq 3.5 \times10^{16}$~cm$^{-2}$ will fall out 
of hydrostatic equilibrium and precipitate onto the disk of the galaxy on 200-Myr timescales. 
Ram-pressure stripping can disrupt infalling clouds, although the trailing fragments may 
re-assemble,  by experiencing lower drag forces than the leading cloud. 

\end{enumerate}
An important question is how long galactic winds last.  Bergvall \etal\ (2016) found a median 
starburst age of 70 Myr in a sample of active star-forming galaxies from the Sloan Digital Sky Survey.
Outflows from superbubbles are driven by supernovae in OB associations and last at least 40~Myr, 
the lifetime of the last star to explode in a coeval starburst.  Some outflows last longer, owing to non-coeval 
star formation in OB associations (Shull \& Saken 1995).  Cloud-wind interaction times ($\sim$50~Myr) 
are comparable to outflow times, but much smaller than timescales for radiative cooling (400-600~Myr) 
and cloud infall (200~Myr).  Disruption of smaller clouds occurs by ram-pressure stripping arising from 
large relative velocities developed during infall or during intermittent galactic outflows.  

Our analytic models of the thermal evolution of cooling clouds made several simplifying approximations, 
such as maintaining spherical shape and constant mass.  We neglected cloud disruption during 
their 200-Myr infall, compared to the 40-70~Myr durations of galactic outflows and cloud interaction.
Cloud disruption through infall has been simulated numerically (e.g., Heitsch \& Putman 2009;  
Armillotta \etal\ 2017) with a consensus that bigger clouds are disrupted more slowly than smaller 
clouds.   In large clouds ($R>250$ pc) a significant fraction of the cloud will survive longer than 250 Myr 
for relative velocities $100-300~\rm{km~s}^{-1}$.   The next steps in our investigation will be to quantify 
the rate at which gas is accreted onto the disk, using numerical techniques to derive the cloud evolution 
($R$, $n$, $T$) with realistic cooling rates and hydrodynamics.  This will enable us to test the assumptions 
of our analytic models.  
We will model the cloud infall, including gravitational and drag forces on the infalling cloud and mass 
loss from interactions with hot halo gas and galactic outflows.  We will also investigate the possibility 
that stripped gas behind the infalling cloud experiences less drag and re-accretes onto the cloud.  
These theoretical studies will explore the parameter-space of cloud evolution, for galactic halos of different 
densities, temperatures, and metallicities.   

\vspace{0.2cm}

\noindent
{\bf Acknowledgements.}
Initial portions of this work were supported by the undergraduate Honors Program at the
University of Colorado Boulder.  We acknowledge helpful discussions with Mark Voit,
Fabian Heitsch, Chris McKee, and Yakov Faerman on astrophysical processes involving 
galactic halo gas.


\small{

}

\end{document}